\documentclass{aa} 
\usepackage{graphicx}
\usepackage{amsmath}
\usepackage[]{natbib}
\usepackage{txfonts}
\usepackage{color}
\begin{document}

  \title{Close binary evolution. III. Impact of tides, wind magnetic braking, and internal angular momentum transport}


  \author{H.F. Song,
     \inst{1,2,3}, G. Meynet\inst{2,\dag}, A. Maeder\inst{2}, S. Ekstr\"om\inst{2}, P. Eggenberger\inst{2}, C. Georgy\inst{2}, Y. Qin\inst{2,4}, T. Fragos\inst{2}, M. Soerensen\inst{2}, F. Barblan\inst{2}, G. A. Wade\inst{5}}
\authorrunning{Song et al.}
\institute{College of Physics, Guizhou University, Guiyang City,
Guizhou Province, 550025, P.R. China \and Geneva Observatory, Geneva
University, CH-1290 Sauverny, Switzerland \and Key Laboratory for
the Structure and Evolution of Celestial Objects, Chinese Academy of
Sciences, Kunming 650011 \and Guangxi Key Laboratory for
Relativistic Astrophysics, Department of Physics, Guangxi
University, Nanning 530004, China \and
Department of Physics, Royal Military College of Canada, Ontario, Canada\\
$^{\dag}$Corresponding author, \email{georges.meynet@unige.ch}
}

  \date{Received; accepted }

 \abstract
 {Massive stars with solar metallicity lose important amounts of rotational angular momentum through their winds. When a magnetic field is present at the surface of a star, efficient angular momentum losses can still be achieved even when the mass-loss rate is very modest, at lower metallicities, or for lower-initial-mass stars. In a close binary system, the effect of wind magnetic braking also interacts with the influence of tides, resulting in a complex evolution of rotation.}
  {We study the interactions between the process of wind magnetic braking and tides in close binary systems.
  }
 {We discuss the evolution of a 10 M$_\odot$ star in a close binary system with a 7 M$_\odot$ companion using the Geneva stellar evolution code. 
 The initial orbital period is 1.2 days.
 The 10 M$_\odot$ star has a surface magnetic field of 1 kG. Various initial rotations are considered.
 We use two different approaches for the internal angular momentum transport. In one of them,  angular momentum is transported by shear and meridional currents. In the other, a strong internal magnetic field imposes nearly perfect solid-body rotation. The evolution of the primary is computed until the first mass-transfer episode occurs. The cases of different values for the magnetic fields and for various orbital periods and mass ratios are briefly discussed. 
  }
 {We show that, independently of the initial rotation rate of the primary and the efficiency of the internal angular momentum transport, the surface rotation of
the primary will converge,
 in a time that is short with respect to the main-sequence lifetime, towards a slowly evolving velocity
 that is different from the synchronization velocity.
This "equilibrium angular velocity" is always inferior to the
angular orbital velocity. In a given close binary system at this
equilibrium stage, the difference between the spin and the orbital
angular velocities becomes larger when the mass losses and/or the
surface magnetic field increase. The treatment of the internal
angular momentum transport has a strong impact on the evolutionary
tracks in the Hertzsprung-Russell Diagram as well as on the changes
of the surface abundances resulting from rotational mixing. Our
modelling suggests that the presence of an undetected close
companion might explain rapidly rotating stars with strong surface
magnetic fields, having ages well above the magnetic braking
timescale. Our models predict that the rotation of most stars of
this type increases as a function of time, except for a
first initial phase in spin-down systems. The measure of their
surface abundances, together, when possible, with their
mass-luminosity ratio, provide interesting constraints on the
transport efficiencies of angular momentum and chemical species.}
 {Close binaries, when studied at phases predating any mass transfer, are key objects to probe the physics of rotation and magnetic fields in stars.
 }

  \maketitle
%

\section{Introduction\label{Sec_intro}}


Many massive stars are in binary systems \citep[see e.g.,][]{Dunstall2015, Moe2016} and a significant fraction of these systems  \citep[about 30\% according to][]{deMink09b}
are tight enough that interactions between the components, either by tides, mass transfer episodes, or merging events, should be expected during their main sequence (MS) lifetimes.
The impact of tidal torques on the evolution of the primary star in
close binaries has been studied recently as a path to trigger strong
tidally induced shear mixing \citep{deMink2009a, Song2013}, and
homogeneous evolution \citep{deMink2009a, Song2016}, especially as
formation mechanism for coalescing binary stellar black holes
\citep{Mandel2016, deMink2016, Marchant2016}.

According to recent large surveys, Magnetism in Massive Stars [MiMeS]
\citep{Wade2016}, and B fields in OB Stars [BOB] \citep{Fossati2015},
about 7-10\% of OB stars show a surface magnetic field above 100 G
\citep{Wade2014}. These magnetic massive stars pose extremely
interesting challenges. One of them is to understand the origin of
such magnetic fields. Although it is generally agreed that these are
quasi-stable remnants of magnetic fields generated earlier in their
evolution \citep[e.g.,][]{Braithwaite2014}, the originating
mechanisms are currently unknown. Another interesting question is
whether such strong surface magnetic fields may impact the evolution
of their host star. For example, surface magnetic fields may
significantly reduce the mass lost through stellar winds
\citep{Petit17}. As shown by these authors, this may allow "heavy" stellar-mass
black holes with masses above $\sim$25 M$_\odot$ to be produced even at solar metallicity where classical
mass loss would prevent such massive black holes from being formed.
Such an effect may also explain the occurrence of Pair Instability
supernovae at solar or higher metallicities \citep{Georgy2017}. A
second important consequence of a surface magnetic field is the wind
magnetic braking effect \citep{UdDoula2002, UdDoula2008}.  By
channeling the matter expelled in the wind to large distances from
the star, magnetic field lines may exert a torque at the surface of
the star. This latter aspect has been explored for massive stars by
\citet{Meynet2011}. These authors showed that, depending on the
efficiency of the angular momentum transport inside the star, very
different surface enrichments in nitrogen are produced. In the
case of solid body rotation, the wind magnetic braking effect
produces slowly rotating MS stars with no surface nitrogen enrichment. In
models for which angular momentum is transported by meridional
currents and shear, wind magnetic braking produces slowly rotating
and strongly nitrogen enriched MS stars.

In the present work we want to address the question of how a massive
star with a strong surface magnetic field would evolve in a close binary system. \citet{Repetto2014}  have already studied such systems but
without following the evolution of the detailed structure of the star as we have done here\footnote{\citet{Eggleton2002} have also already studied
such effects in the context of cool Algol systems.}. They obtain that in certain circumstances{ ``the stellar spin
tends to reach a quasi-equilibrium state, where the effects of tides and winds are counteracting each other.''}. We confirm this interesting effect in the present work.
Since we are following the evolution of the structure of the star, we also study how tides and wind magnetic braking affect the
evolutionary tracks, the changes of the surface composition, and the evolution of the orbit  simultaneously.

We also explore the impact of two different ways of treating the
transport of angular momentum in the interiors of stars. In one
treatment, we apply the theory devised by \citet[][]{Zahn1992} and
\citet[][]{Maeder1998}, where the angular momentum and chemical
species are transported by shear turbulence and meridional currents.
In a second treatment, by introducing a large diffusive coefficient $D_{\Omega}$ in the
advecto-diffusive equation, we consider a very efficient
transport of angular momentum inside the star that would impose
solid-body rotation during the MS phase. In
this last type of model, chemical mixing through shear mostly
disappears but continues to be driven, to a degree, by meridional
currents (see the right panel of Fig.~7).

In Sect. 2, we use an analytic approach to explore the space of initial conditions, allowing tides and magnetic torque to more or less compensate each other
on a timescale shorter than the MS. Numerical models computed using the Geneva stellar evolution code, accounting more consistently for all physical aspects
of the problem than done in the analytic approach, are presented in Sect.~3. Their results are compared with the analytic solutions obtained in Sect.~2, and various consequences
of the interactions of the tidal torques and the wind magnetic braking are discussed for both differentially and solid-body rotating models.
Section~4 synthesizes the main conclusions and proposes some future perspectives.

\section{Tides and magnetic braking; an analytical approach}

\subsection{The equations}

The change of the spin angular momentum (assuming solid-body rotation)
due to tidal interaction is given by \citet{Zahn77}
\begin{equation}
\left ( {dJ \over dt}  \right )_{\rm tide}=-3MR^{2}(\Omega-\omega_{\rm orb})\left ( {GM \over R^{3}} \right )^{1/2}[q^{2}\left ( {R\over a} \right )^{6}]E_{2}s_{22}^{5/3},
\end{equation}
where $J$ is the total angular momentum of the star, $M$, $R$ and
$\Omega$, respectively, are its total mass, radius, and spin angular
velocity, $\omega_{\rm orb}$ is the orbital angular velocity, $a$ is
the separation between the two components of the binary system (we
assume a circular orbit), and $E_2$ is the tidal coefficient that can be
expressed by $E_2\sim 10^{-1.37}(R_{\rm conv}/R)^8$, with $R_{\rm
conv}$ being the radius of the convective core \citep{Yoon2010} and
$s_{22}=2|\Omega-\omega_{\rm orb}| \left (R^3 \over GM\right
)^{1/2}$ \citep{Zahn77}.

In differentially rotating models, $J$ is the angular momentum of the region in the star where the tides deposit or remove angular momentum.
This region is assumed to be the outer layers of the star comprising a few percent of the total mass \citep[see][for more details]{Song2013}.


The angular momentum evolution due to magnetic braking can be
expressed as \citep{UdDoula2002, UdDoula2008}
\begin{equation}
\left ( {dJ \over {\rm d} t} \right )_{\rm mb}\simeq\frac{2}{3}\dot{M}\Omega R^{2}[0.29+(\eta_{\ast}+0.25)^{1/4}]^{2},
\end{equation}
\noindent where $\dot{M}$ is the mass-loss rate,  $\eta_{\ast}=\frac{B_{\rm eq}^2R^2}{\dot{M}\upsilon_\infty}$, with $B_{\rm eq}$ being the equatorial magnetic field, which is equal to
one-half the polar field in the case of a dipolar magnetic field aligned with the rotational axis, and $\upsilon_\infty$  the terminal wind velocity of the wind.



The ratio of the tidal to the magnetic timescale is given by:
\begin{equation}
\frac{\tau_{\rm tides}}{\tau_{\rm mb}}={
 {
 (J/
{dJ \over dt} )
 }_{\rm tide}
   \over
 {
 (J/
{dJ \over d t} )}_{\rm mb} } =\frac { \frac{2}{3}\dot{M} \Omega
R^{2}[0.29+(\eta_{\ast}+0.25)^{1/4}]^{2} } {
3MR^{2}|\Omega-\omega_{\rm orb}|\left (\frac{GM}{R^{3}}\right
)^{1/2}[q^{2}\left (\frac{R}{a}\right )^{6}]E_{2}s_{22}^{5/3} }.
\end{equation}
We have considered here the absolute value of $\Omega-\omega_{\rm orb}$ because timescales
are always positive.


\begin{figure*}
  \centering
  \includegraphics[width=8.9cm]{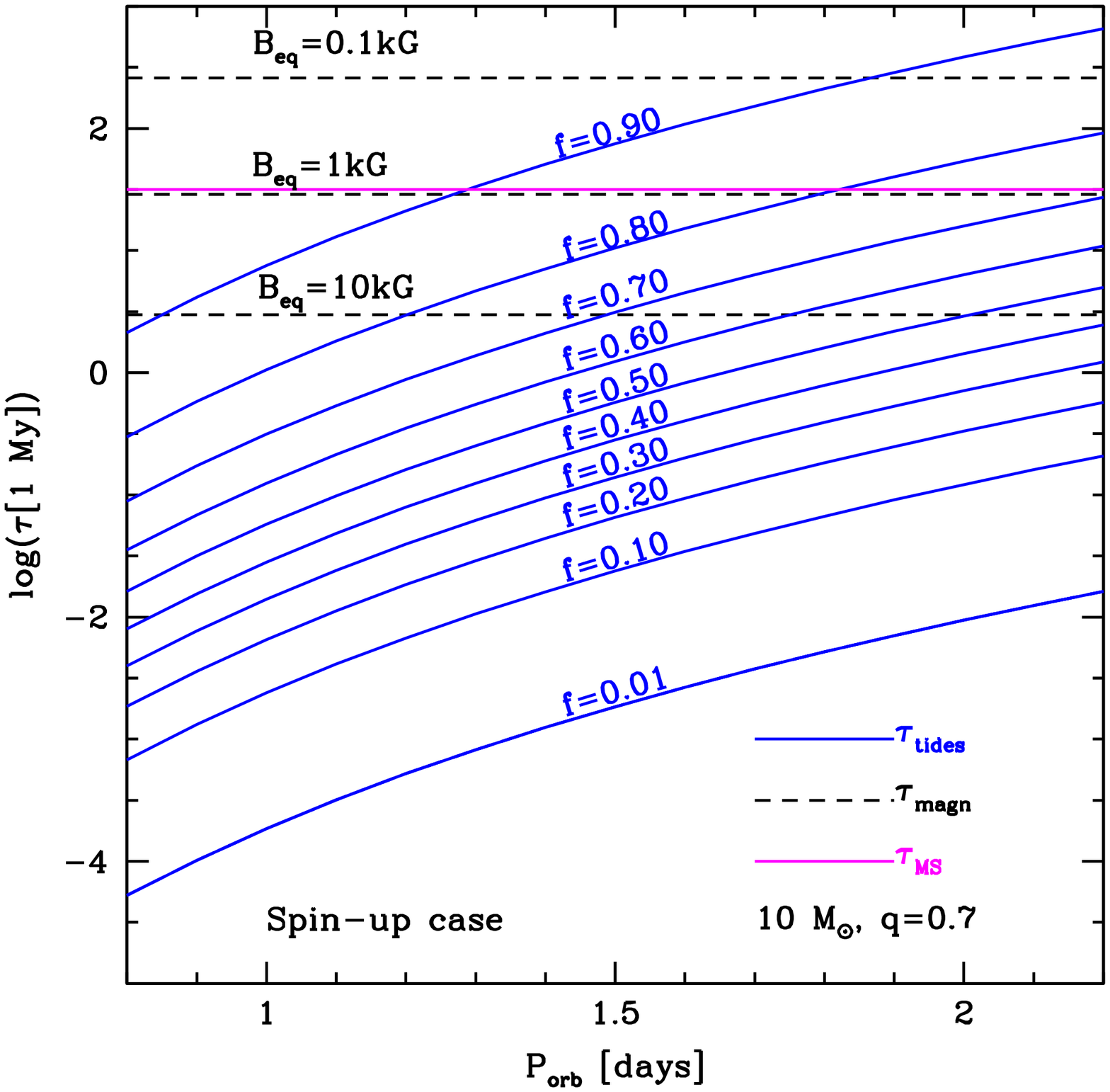}  \includegraphics[width=8.9cm]{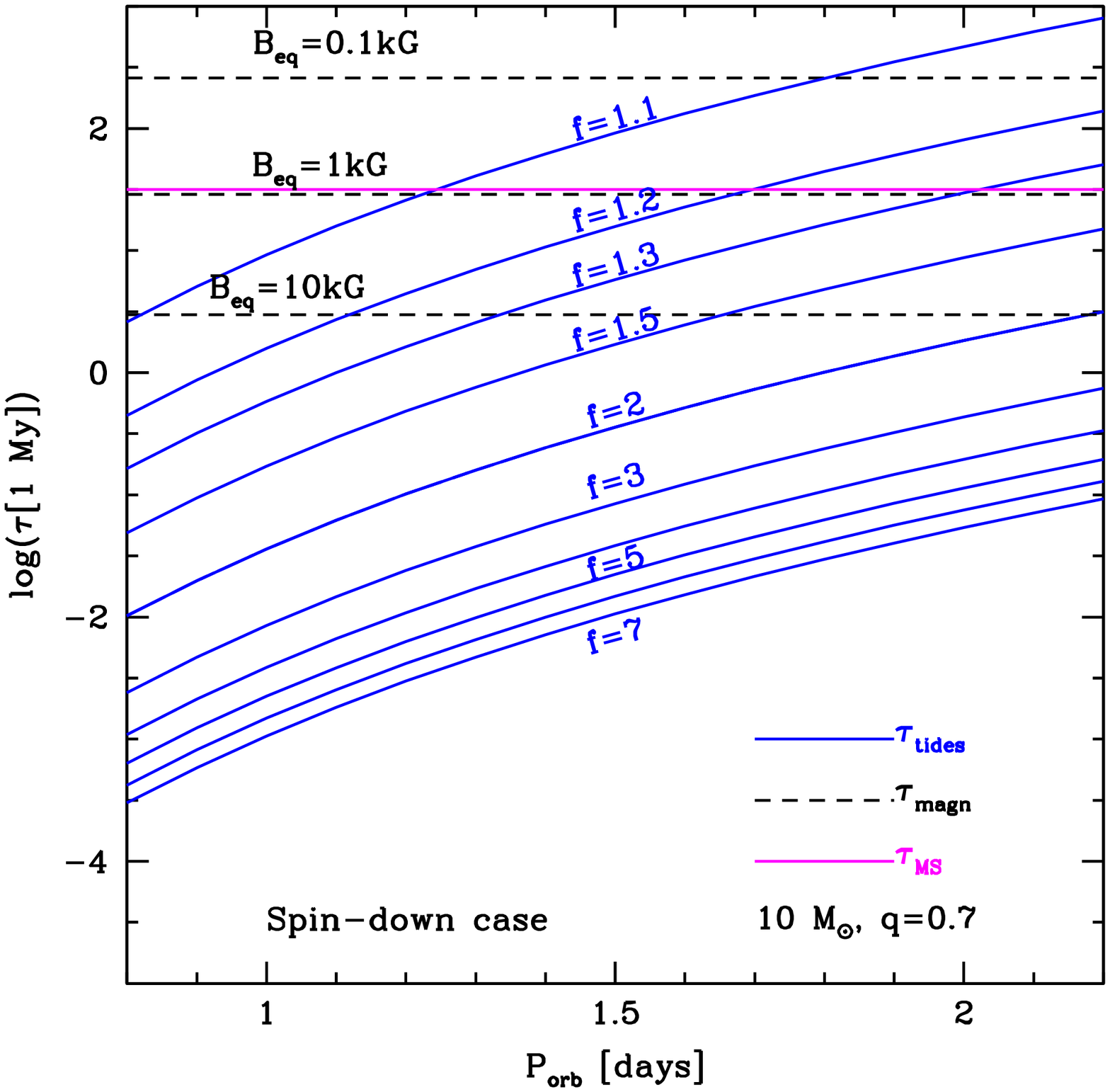}
   \caption{Variation of the timescales for the tidal torques (blue continuous lines)
   and for the wind magnetic braking (black dashed lines) as a function of the orbital period
   for a 10 M$_\odot$ star with Z=0.007. The initial spin angular velocity of the primary is taken to be equal to
   f $\omega_{\rm orb}$.The tidal torque has been estimated assuming that the
   companion is a 7 M$_\odot$ star, that means a mass ratio of the secondary to the primary, $q$, equal to 0.7. The quantities have been estimated adopting zero-age main
   sequence (ZAMS) values for the 10 M$_\odot$ star. The horizontal magenta line
   corresponding to 30 Myr, indicates the typical MS lifetime of a 10 M$_\odot$ star.  {\it Left panel:}
   The case of spin-up by the tidal torque. {\it Right panel:} the case of spin-down.}
     \label{fig0}
\end{figure*}

\subsection{Order-of-magnitude estimates}

Let us consider the case of a binary system with a mass of the
primary $M_1$ equal to 10 M$_\odot$,  and a mass ratio, $q=M_2/M_1$ equal
to 0.7 (thus a mass of secondary, $M_2$, equal to 7 M$_\odot$). We
consider ZAMS values for the mass-loss rate, stellar radius and
radius of the convective core of the 10 M$_\odot$ model. We adopt a
mass-loss rate computed from the relation of \citet{V2001}, with the
terminal velocity, $\upsilon_\infty=2.6\   \upsilon_{\rm esc}$,
given by those same authors (where $\upsilon_{\rm esc}$ is the
escape velocity). In this work, we have accounted for the possible
quenching of the mass-loss rate by the magnetic field as was done by
\citet{Petit17}. Kepler's law is used for linking $a$ to the orbital
period $P_{\rm orb}$. In Eq.~(1), we shall use $P_{\rm orb}$ instead
of the orbital angular velocity ($P_{\rm orb}=2\pi/\omega_{\rm
orb}$). We express the spin angular velocity $\Omega$ at the stellar
surface as a function of $P_{\rm orb}$ introducing a parameter $f$
such that $\Omega= f \omega_{\rm orb}= f 2\pi/P_{\rm
orb}$\footnote{We note that $f$ has an upper limit given by $\Omega
=\Omega_{\rm crit}$, where $\Omega_{\rm crit}$ is the angular
velocity of the primary such that the centrifugal acceleration at
the equator balances the gravity there. The upper limit of $f$ is
$3.1944\ P_{\rm orb}$, with $P_{\rm orb}$ in days. }. Using these
inputs, we obtain

\begin{equation}
\frac{\tau_{\rm tides}}{\tau_{\rm mb}}={
4.948\ \ 10^{34} {f \over P_{\rm orb}} [0.29+(1.537\ \ 10^{4} B_{\rm eq}^2+0.25)^{1/4}]^2
\over
1.0373\ \ 10^{40} {
|f-1|^{8/3} \over P_{\rm orb}^{20/3}}
}
,\end{equation}
where $P_{\rm orb}$ is in days and $B_{\rm eq}$ in kG.

The tidal and magnetic wind torque timescales are plotted in
Fig.~\ref{fig0}. Let us focus on the left panel corresponding to
spin-up cases, and follow the evolution of a star beginning with a
spin angular velocity equal to 1\% the orbital angular velocity
($f=0.01$) and having a 10 kG surface equatorial magnetic field. For
all the orbital periods considered in this plot, the tidal torque
timescale is much shorter than the magnetic torque (see the dashed
horizontal lines) as well as the MS timescale (see the
magenta continuous horizontal line). Thus tides will rapidly spin up
the star, producing an increase of $f$. The star will therefore move
up in this diagram.

The orbital period will also change. On one hand, the system is
losing mass, making the orbit larger and increasing $P_{\rm orb}$. On
the other hand, the tidal torque accelerates the star and thus
transfers angular momentum from the orbit to the star. This tends to
decrease the separation $a$ and therefore $P_{\rm orb}$. 
At a given point however,
the timescale for acceleration by the tides becomes of the same
order of magnitude as the timescale for braking by the magnetic
winds. From this point on, the surface velocity of the star will
remain locked around the value where these two torques compensate
each other. We shall call this velocity the equilibrium velocity.
For an orbital period equal to 1 day, this will correspond to about
85\% of the angular orbital velocity. For an orbital period of 2 days,
this equilibrium velocity will be around 50\% of the angular orbital
velocity. Actually this equilibrium velocity will change since the
separation of the two stars will change as a result of tidal torques
that will continuously transfer angular momentum from the orbit to the star,
compensating for the loss by the winds (magnetic or not).
However, these changes are
slow as will appear in Sect.~3 below.


In the case of spin-down (see the right panel of Fig.~\ref{fig0}), tides will cause the star to move vertically, as previously, since in both cases of spin-up and spin-down, the tidal timescale increases.  Now, in the spin-down case, the magnetic braking will work together
with the tides to slow down the star. At the beginning, tides are stronger than the magnetic torque. However the strength of the tides decreases while that of the magnetic braking remains more or less constant (we note that here
it remains strictly constant because we assumed ZAMS values for all the ingredients entering the wind magnetic braking timescale).
When the tidal torque becomes zero ({\it i.e.,} when $\Omega$ = $\omega_{\rm orb}$), the magnetic braking is still active and will cause $\Omega$  to overshoot below $\omega_{\rm orb}$, until
a point when the tidal torque becomes again larger than the wind magnetic braking. But this time, since $\Omega$ will be below
$\omega_{\rm orb}$, tides will accelerate the star and thus counteract the braking due to the winds.  From this stage on
we shall have a situation similar to that discussed in the spin-up case. The star will finally settle around an equilibrium velocity that will be similar to the one reached in the spin-up case.

The equilibrium velocity depends on many factors (as the masses and separation of the two components, the tidal forces, the mass loss rates, etc.) and in particular on the strength of the surface magnetic field. A strong field causes the rotation of the star to approach a smaller  equilibrium velocity than a weak field.
For a quite unrealistic value, $B_{\rm eq}$=1000 kG, the equilibrium velocities would be between a few percent  and 45\% of the orbital velocity
for $P_{\rm orb}$ =2 and 1 day,  respectively.
In the case where $B_{\rm eq}= $ 1 kG, tides will provide the main torque since the magnetic
braking timescale is of the same order of magnitude as
the MS timescale.
The surface velocity of the star will converge towards values above 90\% of the orbital velocity at
$P_{\rm orb}$= 1 day and around 70\% of the orbital velocity at
$P_{\rm orb}$= 2 days.

\begin{figure}
  \centering
  \includegraphics[width=8.5cm]{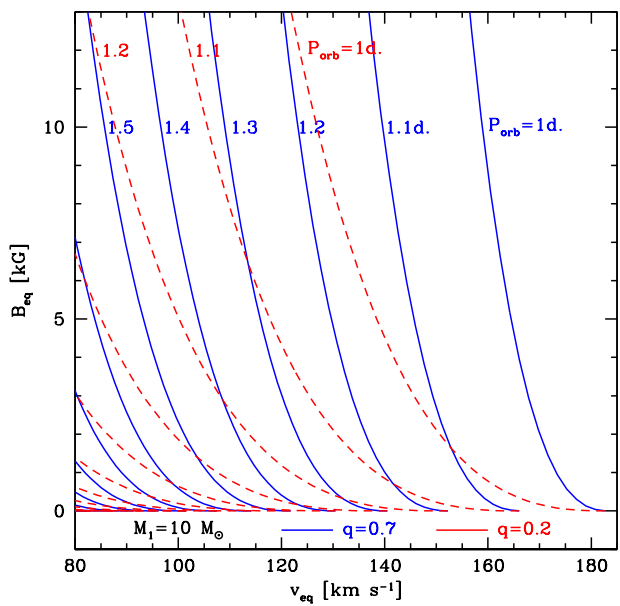}
   \caption{This plot allows determination of the rotation velocity of the primary (x-axis) when the equilibrium state is reached ($\frac{\tau_{\rm tides}}{\tau_{\rm mb}}=1$) for
   various values of the surface magnetic field (y-axis). The relations between these two quantities are given by the lines. For example,
   the rotation of the primary in a system composed of a 10 and a 7 M$_\odot$ star (blue continuous lines) when the orbital period is 1.1 days, assuming the primary
   has a surface magnetic field of 10 kG, would be around 137 km s$^{-1}$. 
   The blue continuous lines are for a mass ratio $q=M_2/M_1$=0.7
 and the red dashed lines are for a mass ratio $q=0.2$.}
     \label{fig1}
\end{figure}

In Fig.~\ref{fig1}, we have plotted lines corresponding to $\frac{\tau_{\rm tides}}{\tau_{\rm mb}}=1$ in the plane surface (equatorial) magnetic field  versus stellar equilibrium rotation velocities at the equator. We see that
a 10 M$_\odot$ primary star with an equatorial surface magnetic field of 10 kG, in a binary system with a 7 M$_\odot$ companion and an orbital period equal to 1 day, will reach, after the convergence process, a surface velocity of about 160 km s$^{-1}$.

When the orbital period increases, rotation will converge towards lower values since the tidal torque is smaller. The same is true when the mass ratio decreases\footnote{For a general $q$ value, Eq.~(4) has to be multiplied by a factor equal to
0.215 $(10q(1+q))^{1/3}/q^2$.}. We note that there is no difference in Fig.2 whether a spin-up or a spin-down is considered. Indeed, as mentioned above, in case of spin-down, the magnetic braking will, at a given time, overcome the tidal torque until the surface rotation decreases to a value such that tidal torque will
again accelerate the star. Thus we shall be back to a situation of spin-up. The numerical models that are discussed
in the following section will show however that the two cases of spin-up and spin-down are not completely equivalent.

\section{Tides and magnetic braking; a numerical approach}

The evolution of a 10 M$_\odot$ stellar model in a close binary system with a 7 M$_\odot$ companion is computed with the Geneva stellar evolution code.
The 10 M$_\odot$ star has a surface magnetic field of 1 kG. We assume this value to be constant during the MS phase.

An initial orbital period equal to 1.2 days was considered \footnote{For a system of mass $M_{1}+M_{2}=10 M_{\odot}+7 M_{\odot}$, one has a semi-major axis $a=7.52\ \ 10^{11} P_{\rm orb}^{2/3}$ (in units of cm) where $P_{\rm orb}$ is in days.}.
Two different initial rotation rates for the 10 M$_\odot$ model are considered: a value $\upsilon_{\rm ini}$= 310 km s$^{-1}$ for the spin-down cases and $\upsilon_{\rm ini}$= 60 km s$^{-1}$ for the spin-up cases.
We assume the secondary star is a point mass with negligible mass loss during the period considered.

We used two different prescriptions for the transport of angular
momentum inside the star. We considered the case where a moderately
efficient transport occurs, allowing a moderate radial differential
rotation to develop inside the star. This case corresponds to the
shellular rotation model proposed by \citet{Zahn1992, Maeder1998}.
The same prescriptions for rotation described by \citet{Ekstrom2012}
are used to compute these models. In the following we label these
models using the capital letter D (for `differential rotation').

We also consider the case of solid body rotation. This case corresponds to the case of very efficient internal angular momentum transport.
For instance, in these models any change of angular momentum at the stellar surface has an instantaneous impact on the angular momentum distribution in the whole star. For computing these models
we used the same prescriptions for rotation as described by \citet{Song2016}\footnote{This very efficient transport is mediated in the present models through the
theory devised by \citet{Spruit1999, Spruit2002} that postulates the activity of an efficient dynamo in differentially rotating radiative layers of the stars. Actually for our purpose, the mechanism responsible for
this efficient angular momentum transport is not very important provided that it is efficient enough to impose solid body rotation \citep[see also][]{Song2016}.}.
In the following, we label these models with an S (for `solid-body rotation').
In these models, the mixing of the chemical elements is driven by meridional currents.

\subsection{The orbital evolution}


In our numerical approach, we account for the evolution of the orbit as a function of time in the following way.
The orbital angular momentum of the system is $J_{\rm orb}=M_{1}M_{2}(\frac{Ga}{M_{1}+M_{2}})^{1/2}$ where $a$ is the orbital separation assuming a circular orbit\footnote{This expression implicitly assumes that there is no wind material retained by the system and participating in the orbital evolution. One assumes the winds are sufficiently fast to not feel any significant gravitational attraction when flowing out of the system.}. The variation of the orbital separation is given by
taking the time derivative of $J_{\rm orb}$. We deduce that
\begin{equation}
\frac{\dot{a}}{a}=2\frac{\dot{J}_{\rm orb}}{J_{\rm orb}}-2\frac{\dot{M}_{\rm 1,wind}}{M_{1}}-2\frac{\dot{M}_{\rm 2, wind}}{M_{2}}+\frac{\dot{M}_{\rm 1,wind}+\dot{M}_{\rm 2,wind}}{M_{1}+M_{2}},
\end{equation}
\noindent where mass-loss rates due to stellar winds for the two components are $\dot{M}_{\rm 1,wind}$ and $\dot{M}_{\rm 2,wind}$, respectively. The secondary star is treated as a point mass and its stellar wind has been neglected ($\dot{M}_{\rm 2, wind}=0.0$). Therefore, the above equation becomes
\begin{equation}
\frac{\dot{a}}{a}=2\frac{\dot{J}_{\rm orb}}{J_{\rm orb}}-2\frac{\dot{M}_{\rm 1,wind}}{M_{1}}+\frac{\dot{M}_{\rm 1,wind}}{M_{1}+M_{2}}.
\end{equation}

The variation of the orbital angular momentum is given by $\dot{J}_{orb}=-\dot{J}_{tide} + \dot{J}_{wind}$, where $\dot{J}_{tide}$ is given by Eq.~(1), and
$\dot{J}_{\rm wind}= \dot{M}_{\rm 1,wind} (\frac{M_{2}}{M_{1}+M_{2}}a)^{2}\omega_{\rm orb}$.  This last quantity accounts for the fact that the matter outflowing from the primary star initially has an angular velocity equal to the
orbital velocity of the star\footnote{Actually the wind would also have an angular momentum originating from the spin of the star, but we are only interested here in the {\it orbital angular momentum} that is lost by the winds. We note that this term is identical to the last term in Eqs.~7 and 11 of \citet{Eggleton2002}.}.
The quantity $(\frac{M_{2}}{M_{1}+M_{2}}a)^{2}\omega_{\rm orb}$  is the specific orbital angular momentum of the primary star.

At first sight, it might appear strange that the wind magnetic braking does not appear explicitly in this expression. Actually, $a$ depends on the orbital angular momentum content, not on the spin angular momentum of the stars.
Any effect that changes the spins of the star (such as the mass loss with or without magnetic braking) is accounted for through $\dot{J}_{tide}$ (that scales with $\Omega-\omega_{\rm orb}$).

Expression (6) can be simplified, replacing $\omega_{\rm orb}$ by $[\frac{G(M_{1}+M_{2})}{a^{3}}]^{1/2}$, and using the above expression for $\dot{J}_{\rm wind}$,
\begin{equation}
\frac{\dot{a}}{a}=-2\frac{\dot{J}_{\rm tide} }{J_{\rm orb}}-\frac{\dot{M}_{\rm 1,wind}}{M_{1}+M_{2}}.
\end{equation}

\begin{figure*}
   \centering
\includegraphics[width=8.5cm]{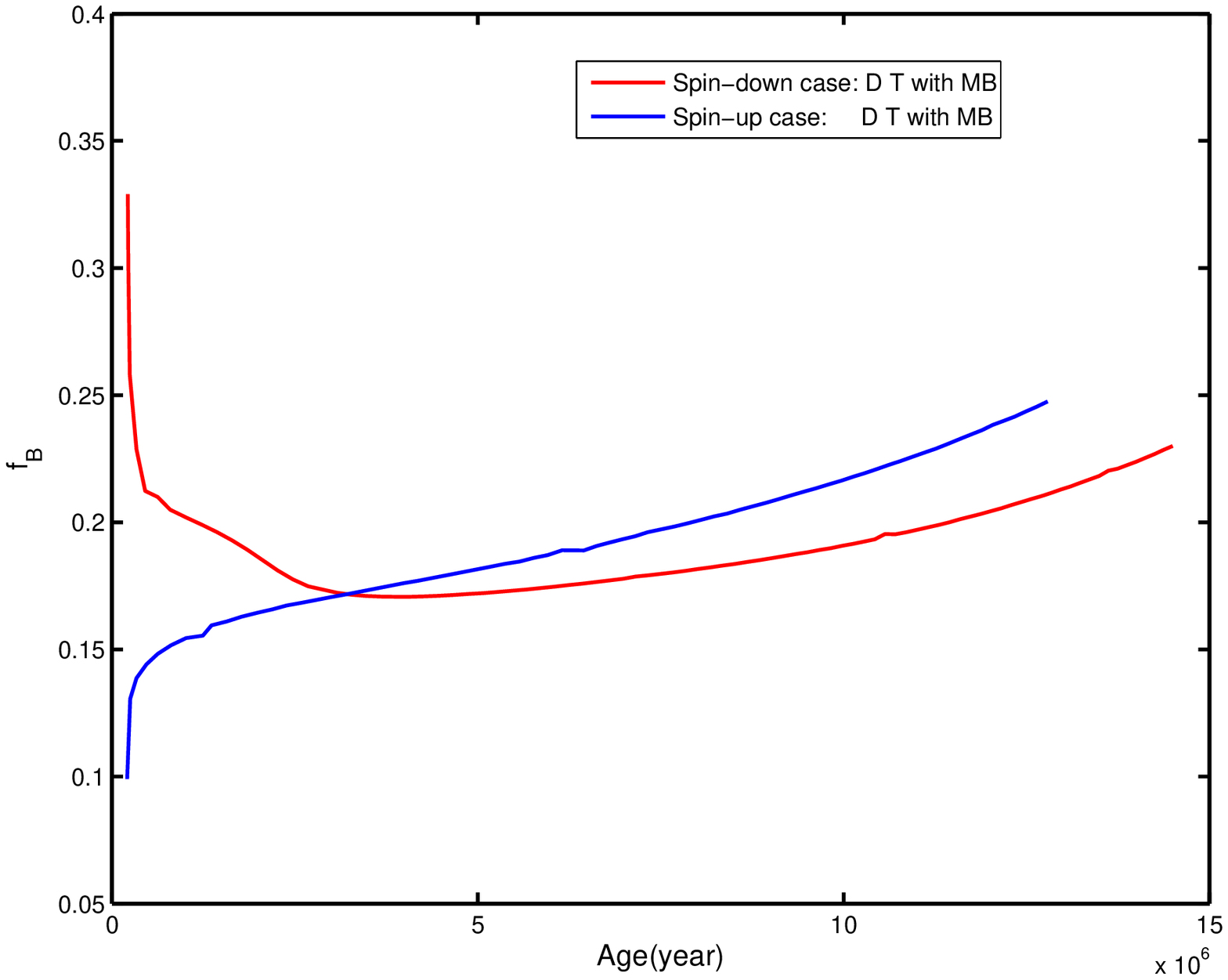}\includegraphics[width=8.5cm]{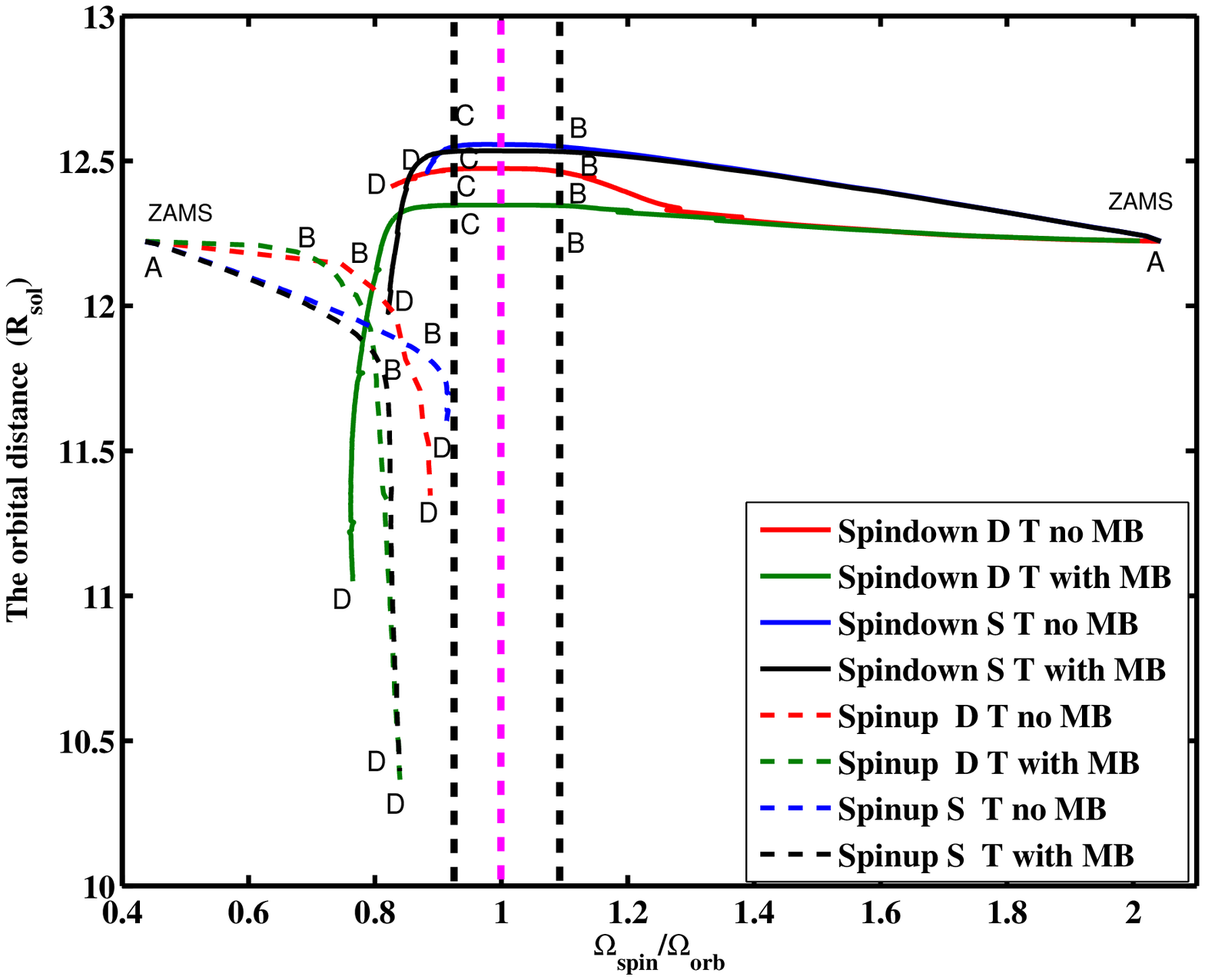}
\caption{{\it Left:} Evolution of the ratio, $f_{\rm B}$ between $\dot{M}_{\rm 1,wind}$ and the mass-loss rate without a magnetic field $\dot{M}_{\rm 1,wind}(B=0)$ for spin-down and spin-up cases with differential (D) rotation, tidal forces (T) and wind magnetic braking (MB).  {\it Right:} Evolution of the separation, expressed in units of solar radii, for spin-down and spin-up cases with both solid (S) body and differential (D) rotation as a function of the spin angular velocity of the primary normalized to the orbital angular velocity. Point A denotes the ZAMS. Tidal torques dominate the orbital evolution from point A to point B in both spin-down and spin-up cases ($\tau_{\rm tides} < \tau_{\rm mb}$).  At point B, one has that $\tau_{\rm tides}=\tau_{\rm mb}$ The loss of mass due to stellar winds governs the orbital evolution from point B to point C in spin-down cases  ($\tau_{\rm tides} > \tau_{\rm mb}$). The wind torques (or magnetic torques) are counteracted by tidal torques from point C to D in spin-down cases and from point B to D in spin-up cases. Point D denotes the beginning of Roche lobe overflow. The red and the green curves show the cases where the redistribution of angular momentum inside the star is due to shear instabilities and meridional currents (cases noted D in the box). The blue and the black curves are cases of solid body rotation (cases noted S in the box)  R$_{\rm sun}$ is the solar radius. The dashed vertical lines correspond to perfect synchronization ($\Omega=\omega_{\rm orb}$ (magenta line) and to 90\% and 110\% of the orbital velocity (black lines left and right of magenta line, respectively).}
         \label{FigVibStab}
   \end{figure*}

For computing $\dot{M}_{\rm 1,wind}$, we have accounted for the fact that, as discussed by \citet{Petit17}, the mass-loss rate is significantly
reduced when a sufficiently strong surface magnetic field is present.  The factor  $f_{\rm B}$ equal to $\dot M_{\rm 1,wind}/\dot M_{\rm 1,wind}$(B=0) is shown in the
left panel of Fig.~\ref{FigVibStab}. We see that the reduction of the mass-loss rate with respect to the non-magnetic case is significant. On the other hand, as explained further below,
this effect should not be very critical for what concerns the results obtained in the present work ({\it i.e.,} ignoring this reduction factor does
not significantly change the results of the present paper.)

Another effect that might impact the results is the gravitational influence of the secondary. Indeed, the gravity exerted by the companion changes the orbital angular momentum lost by the wind.
For example, some material can be accreted by the secondary, or can be trapped in its gravitational potential well. The secondary can also cause some wind material to have its
initial orbital angular momentum decreased or boosted.
This issue has been discussed, for example, by \citet{Brook1993}. 
They found that the orbital angular momentum that is lost is larger or smaller than that inferred neglecting the presence of the secondary
depending on many parameters (such as the mass ratio, the geometry of the surface where matter
is injected into the wind, and the injection velocity of wind particles).  The average angular momentum of escaped particles may change by factors between 0.5 and 20
(see Table 1 in the above reference). Let us suppose that such an effect would change the loss
of orbital angular momentum by a factor 10. In our Eq.~7 above, such an effect would multiply the term dependent upon the mass-loss rate by a factor $\sim$12. We note that this factor
is valid only for the cases with no magnetic field, since the work by  \citet{Brook1993} does not consider this effect. 

Would  such  a change have strong consequences for the evolution of the separation? 
The effect would be limited for the following reason: In Eq.~7, the term $-2\frac{\dot{J}_{\rm tide} }{J_{\rm orb}}$ is in general significantly larger than the term $\frac{\dot{M}_{\rm 1,wind}}{M_{1}+M_{2}}$. Typically,
for a difference $|f-1|$ of 0.1  between the spin and orbital angular velocity (a typical value at the equilibrium stage),  and for an orbital period of 1 day, the first term is about 260 times larger than the second one.
Increasing the mass loss rate term by a factor 12 would keep the first term more than 22 times larger than the second one, and thus tides would remain the dominant term for the evolution of the separation.

Now let us discuss the case with a surface magnetic field attached to the primary as studied here\footnote{We note that
we assume here that the secondary has no strong surface magnetic field.}.  In case of magnetic wind, the wind material will be less sensitive to the gravitational attraction of the companion than without magnetic field.  Indeed if we approximate the evolution of the wind velocity by a $\beta$-law \citep[see the book by][]{Lamers99} then at a distance of about 6 stellar radii, the wind has nearly reached the escape velocity from the primary, namely a value around 1000 km s$^{-1}$.  The ratio of the kinetic energy to the gravitational potential of the secondary at that point is equal to 2.3. The ratio between the magnetic energy and the kinetic energy is 3460, assuming that the magnetic field at the surface of the star is 1 kG and,
that for a dipolar magnetic field, the magnetic field decreases with the distance $r$ as $1/r^3$.
Thus magnetism is largely dominant and likely the gravitational field of the secondary can be neglected. 

There is however another effect that enters into the game when a strong magnetic field is considered: 
the matter launched into the magnetized wind will keep the orbital angular velocity of the primary star until it reaches the Alv\'en radius. The matter launched into the equatorial plan may thus carry away an angular momentum
that is increased by a factor
very roughly equal to the square of the ratio between the Alfv\'en radius ($\sim$ 40 R$_\odot$) and the distance to the center of mass $\frac{M_{2}}{M_{1}+M_{2}}a$ ($\sim$ 6 R$_\odot$) with respect to the mass-loss rate term considered in
Eq. (7).  This factor is equal to about $(40/6)^2=44.4$. This is an upper limit because not all the mass is launched into the equatorial plane. Along the polar axis for instance,
the expression used in the present work would apply. Again, the tidal term would still remain the dominant term, being 260/44.4 $\sim$ 6 times larger than the mass-loss rate term. Therefore this effect would again have rather limited consequence.

Fig.~\ref{FigVibStab} shows the variation of the orbital separation as a function of the ratio of the spin angular velocity to the orbital angular velocity for the primary star. The evolution along the horizontal axis is determined by many different processes.
The spin surface velocity of the primary star depends on its mass-loss rate, the strength of the surface magnetic field, on the change of the stellar radius, and of the angular momentum transport inside the star. The orbital velocity is linked to changes of the orbital angular momentum through mass loss and redistribution of angular momentum between the primary star and the orbit through tides. The evolution along the vertical axis reflects the exchange of angular momentum between the star and the orbit through tides and the impact of the mass lost by the primary (see Eq.~7).

We see that in case of spin-up, the orbital separation always decreases, while in the case of spin-down, the orbital separation
first increases and then decreases.

In the case of spin-up, tides always counteract the wind magnetic braking process, and thus there is continuous transfer of angular momentum from the orbit to the star. This effect dominates the $\dot{M}_{\rm 1,wind}$ in Eq.~7, and thus the separation reduces.
In the case of spin-down, tides and mass loss with or without magnetic braking begin to act together, spinning down the star. The tides transfer angular momentum from the star to the orbit and thus the separation increases.
As already noted above, at a given point, the tidal torque becomes very small or - at synchronization - even disappears. On the other hand, the wind (possibly magnetic)
braking process remains.
Thus $\Omega$ evolves below $\omega_{\rm orb}$, until tidal acceleration becomes strong enough to keep $\Omega$ at a given value that will evolve slowly with time (this happens when the net torque resulting from both
the wind braking and the tides has a timescale that becomes longer than the evolutionary timescale). From this point on, the separation decreases. 

Most of the tracks in Fig.~3 end with a nearly vertical drop that corresponds to the phase when the spin angular velocity of the primary star has converged towards the near-equilibrium velocity. We see that this equilibrium velocity is around 80\% of the orbital velocity
for models with magnetic braking. We see that when magnetic braking is accounted for, the variations of the orbital separation are larger than in the non-magnetic case. This is because more angular momentum is transferred from the orbit to the star when the star loses its angular momentum more efficiently.



\subsection{The evolution of the surface velocities}

\begin{figure*}
  \centering
  \includegraphics[width=8.9cm]{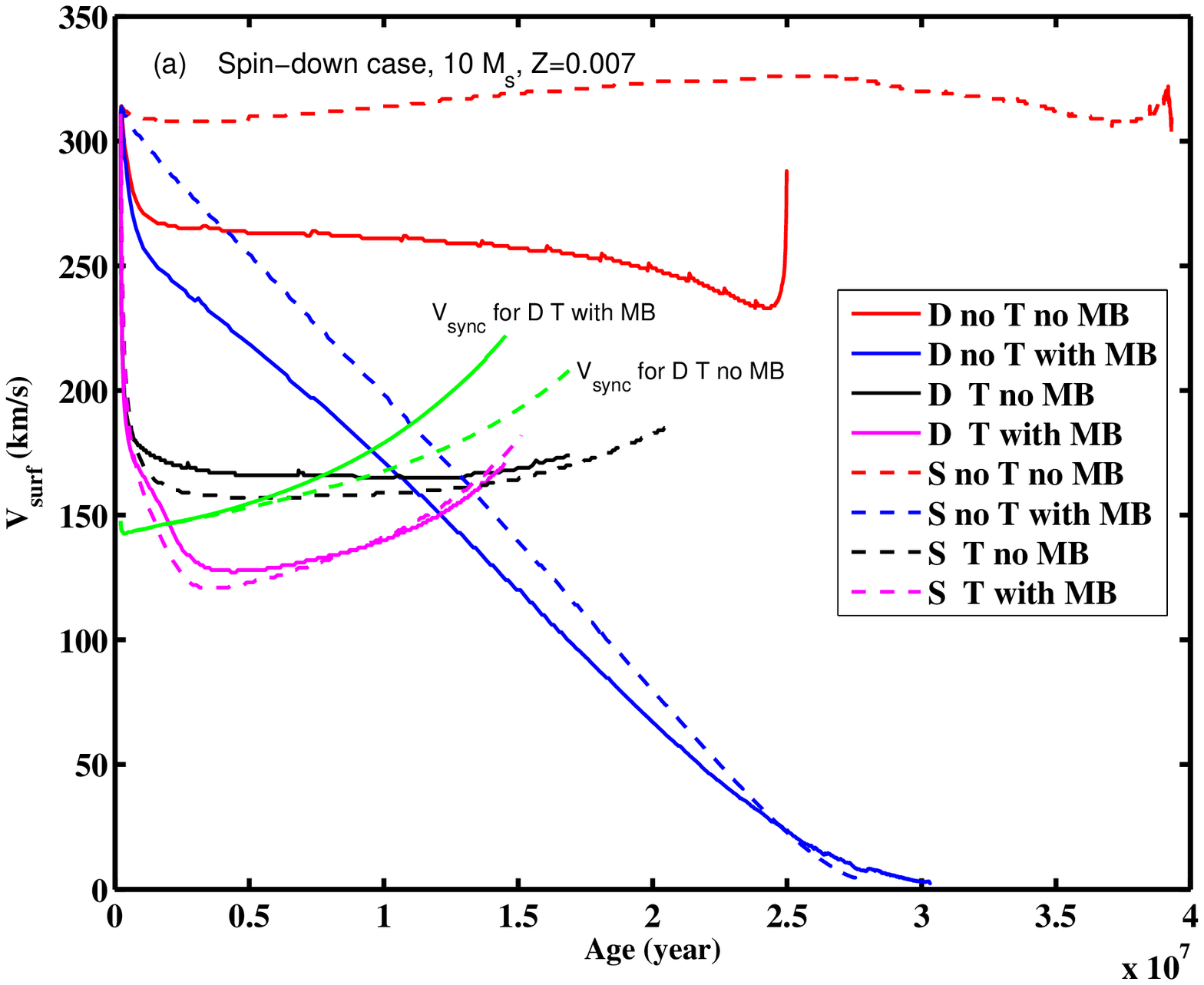}\hfill\hfill \includegraphics[width=8.9cm]{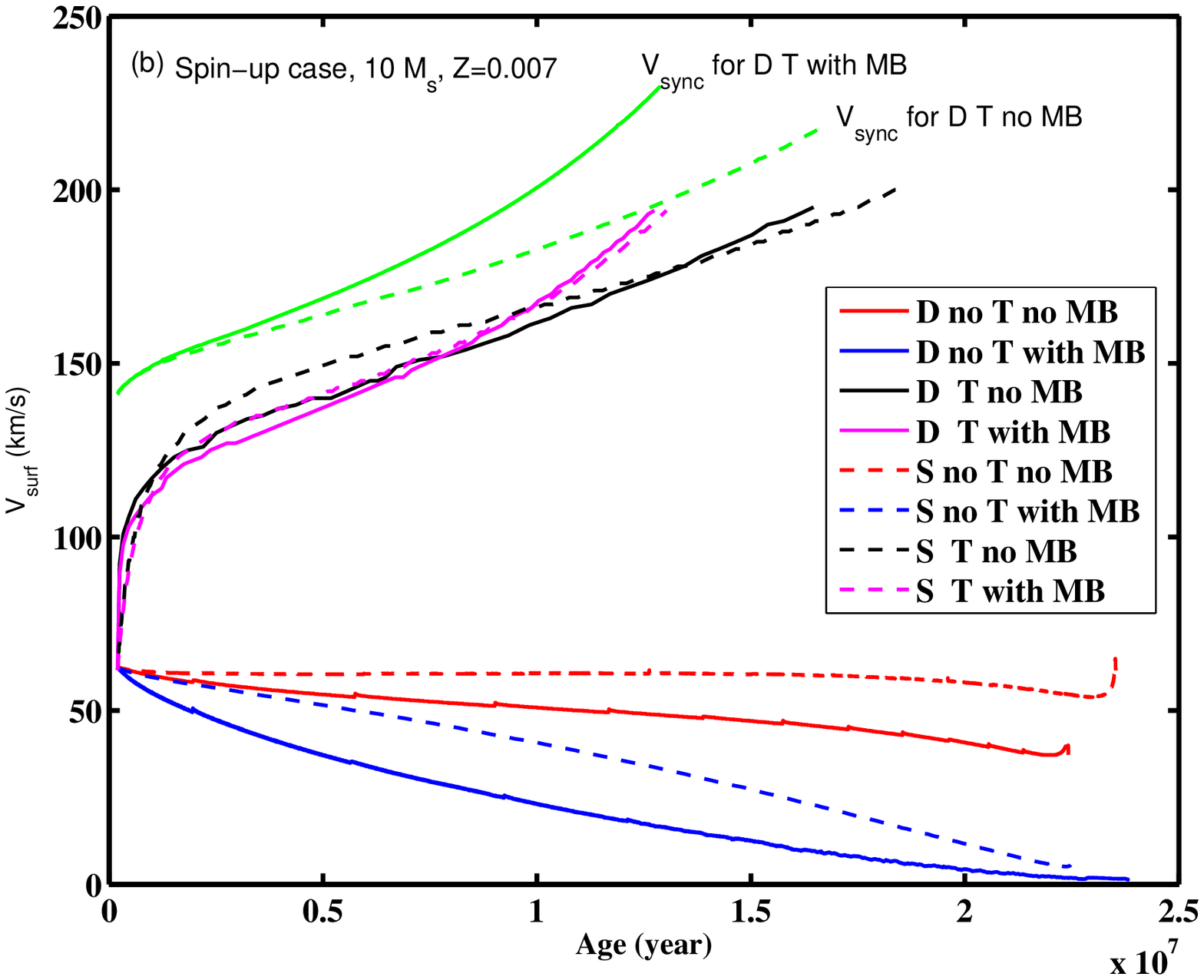}
   \caption{Evolution of the surface equatorial velocity as a function of time for different 10 M$_\odot$ stellar models at a metallicity $Z$=0.007. The continuous curves show the results when the internal angular momentum transport mechanisms
are moderate and allow differential (D) rotation to set up inside the star. The dashed lines show the situation when, due to a very efficient angular momentum transport mechanism, the stars rotate nearly as solid bodies (hence the S labeling the dashed curves). The red curves show the evolution in the case where the star is single (`no T' means `no tidal interaction') and with no magnetic braking (`no MB'). The continuous blue curve shows the situation when magnetic braking is accounted for assuming a constant equatorial surface magnetic field of 1 kG. The black and magenta continuous curves are for the same 10~$M_\odot$ star, now as the primary star in a close binary system with a companion of 7 M$_\odot$ and an initial orbital period of 1.2 days. The black and magenta curves are for models without and with magnetic braking, respectively. The green curves indicate the evolution of the synchronization velocity for the differentially rotating models (the cases for solid rotation are not shown so as to not overcrowd the picture). 
{\it Left panel:} The star begins its evolution with a rotation on the ZAMS equal to 310 km s$^{-1}$. In the close binary considered, the star is being spun down by the tidal forces.
   {\it Right panel:} The star begins its evolution with a rotation on the ZAMS equal to 60 km s$^{-1}$. In the close binary considered, the star is being spun up by the tidal forces.
   }
     \label{v12}
  \end{figure*}

The evolutions of the surface velocities for various 10 M$_\odot$ primary stars are shown in Fig.~\ref{v12}. The two red curves correspond to the evolution of stars with no interaction with a companion (either because the star is single or in a wide binary system). The model allowing an internal radial differential rotation has a lower surface velocity than the solid-body rotating model.  This reflects two processes. First, starting from solid-body rotating ZAMS models,
strong meridional currents rapidly remove angular momentum from the outer layers and transport it toward the central regions until some equilibrium situation sets in between this inward transport and the outward transport by the shear \citep{Deni1999, MM2000}. Secondly, once this large redistribution is terminated, meridional currents change sign and begin to transport angular momentum from the central regions toward the surface. This compensates somewhat for the loss of angular momentum via the stellar wind.

When solid body rotation is considered, the internal redistribution of angular momentum described immediately above obviously cannot occur and thus there is 
no initial strong decrease of the surface velocity.

Comparing the left panel (high initial velocity) to the right panel (moderate initial rotation) of Fig.4, we see that at low velocities the initial drop of the surface velocity in differentially rotating models is much more modest than at high velocities. This is because meridional currents approximately scale with $\Omega$.

Now let us investigate what happens when the rotation of a single star is slowed by the wind magnetic braking mechanism (blue curves in Fig. 4). We see that in all cases, the surface velocity rapidly decreases. The decrease is relatively strong for the fast spinning models and more gentle in the cases of the initially slowly rotating models. This illustrates the fact that a moderate equatorial surface magnetic field (1 kG) can prevent a 10 M$_\odot$ star from being a fast rotator at the end of the MS phase.

When a sufficiently nearby companion is present, tidal forces may spin down or spin up the 10 M$_\odot$ primary star, depending on its initial angular velocity. In Fig. 4, the cases with no magnetic braking are shown by the black curves, those with magnetic braking by the magenta curves. A few interesting points may be noted:
\begin{itemize}
\item Whatever the strength of the surface magnetic field between 0 and 1 kG, and whatever choice has been made between the two options for the internal angular momentum transport, after a rapid transition due to tides, the surface velocity
evolves slowly. The beginning of the slow phase is characterized by a surface equatorial velocity of about 170 km s$^{-1}$ in the case of spin down, and of about 120 km s$^{-1}$ in the case of spin up.
\item In the case of spin down, the velocity curves are convex, showing a minimum surface velocity. In the case of spin up, we see a continual increase of the surface velocity as a function of time.
\end{itemize}

Let us now try to understand these features. The first point reflects the fact that, as expected from timescale arguments presented in Sect.~2, tides are initially the dominant effect. The beginning of the slow phase would correspond to what we call the equilibrium stage in Sect.~2, {that is,} the stage at which the tidal torque becomes
comparable to the wind magnetic torque. Looking at Fig.~\ref{fig1}, we see that for a value of $B_{\rm eq}$ = 1 kG and an orbital period of 1.2 days, the
equilibrium velocity is 145 km s$^{-1}$ , which is between the velocities quoted above at the beginning of the slow phase, namely 120 and 170 km s$^{-1}$ for the
spin down and spin up cases, respectively. Let us stress here that to obtain Fig.~\ref{fig1}, we assumed that the star retains the characteristics it had on the ZAMS all along its evolution, which is obviously a rough assumption. Despite this, we obtain a value comparable to that deduced from the more detailed numerical simulations. The difference of velocities between the spin-down and spin-up cases comes from the fact that these two situations are not
exactly symmetric. This is also apparent looking at the different shapes of the curves underlined in the second point mentioned above.

A first obvious difference between the spin-up and spin-down cases is of course due to the fact that, depending on the initial rotation rate
of the primary, its initial angular momentum is different. Stellar winds (magnetic or not) remove angular momentum from the surface. Internal redistribution of angular momentum will then tap
from the internal angular momentum of the star to compensate for these losses at the surface (in general angular momentum is transported from the inner regions to the envelope either by the transport due to magnetic fields
in the solid-body rotating models, or by meridional currents in the differentially rotating models). Thus, depending on the initial spin angular momentum content of the star, one can expect differences in the outcomes. 

Another effect is that in the case of spin down, and only in the case of spin down, the synchronization stage is reached before the near-equilibrium state.  At synchronization, only the magnetic braking is active.
Thus the equilibrium velocity is approached from a situation where the braking (due to winds, magnetic or not)
is larger than the tides. When the timescale for the net braking torque, that is, resulting from the difference between the wind and tidal torques, becomes sufficiently large, the spin velocity decreases slowly.
This explains that in the case of spin down, the curves show a minimum value.
Why do they increase thereafter?
Since angular momentum is continuously transferred from the orbit to the star, the separation decreases. This reinforces the tides that eventually
accelerate the star and thus the velocity will slowly increase after that minimum.

In contrast, when we have a situation of spin up, the surface equatorial velocity is always increasing.
This can be explained by the fact that at the beginning, the accelerating tides dominate the magnetic braking. After, tides are 
evolving as a result of two counteracting effects. On one side, 
tides decrease due to the fact the $\Omega$ approaches $\omega_{\rm orb}$. On the other side, 
the separation decreases, increasing the tides. This last effect always dominates over the first one, hence
the surface velocity is always increasing, preventing a perfect synchronization.

From the discussion above we conclude that only spin-down cases show a phase during which the stars are deaccelerated, otherwise stars are always accelerated, even if they possess a strong surface magnetic field.
This also implies that the possibility to measure the time variation of the spin periods of stars would be an interesting way to distinguish between the spin-down and spin-up cases, provided of course that
the system corresponds to the initial period where the difference arises.

We see also that the binary systems in Fig.~4 evolve very similarly for the differentially and solid body rotating cases. Thus changing between these two transport mechanisms has little influence on the evolution of the surface velocities.

\subsection{The evolution in the HR diagram}

\begin{figure*}
  \centering
  \includegraphics[width=8.9cm]{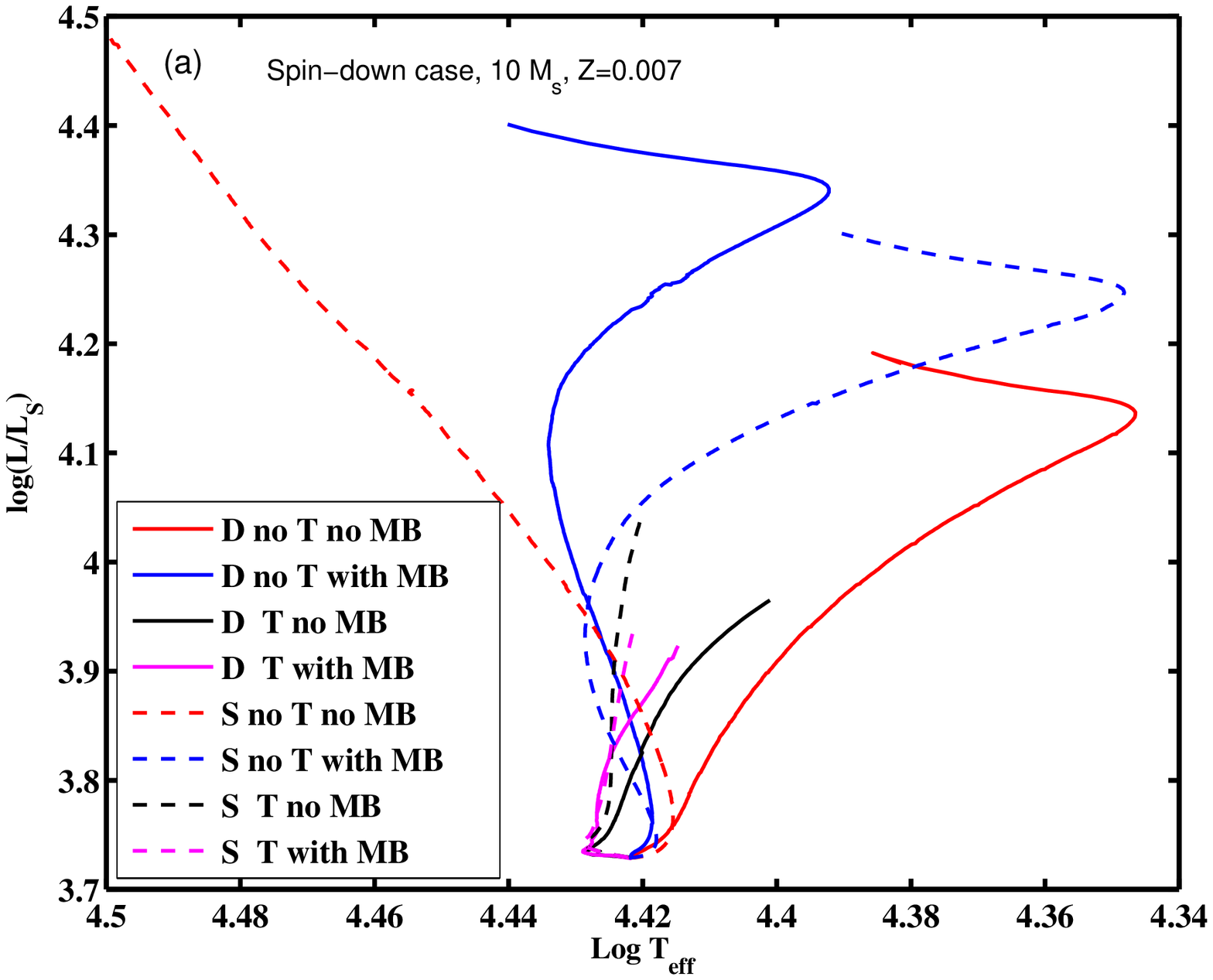}\hfill\hfill \includegraphics[width=8.9cm]{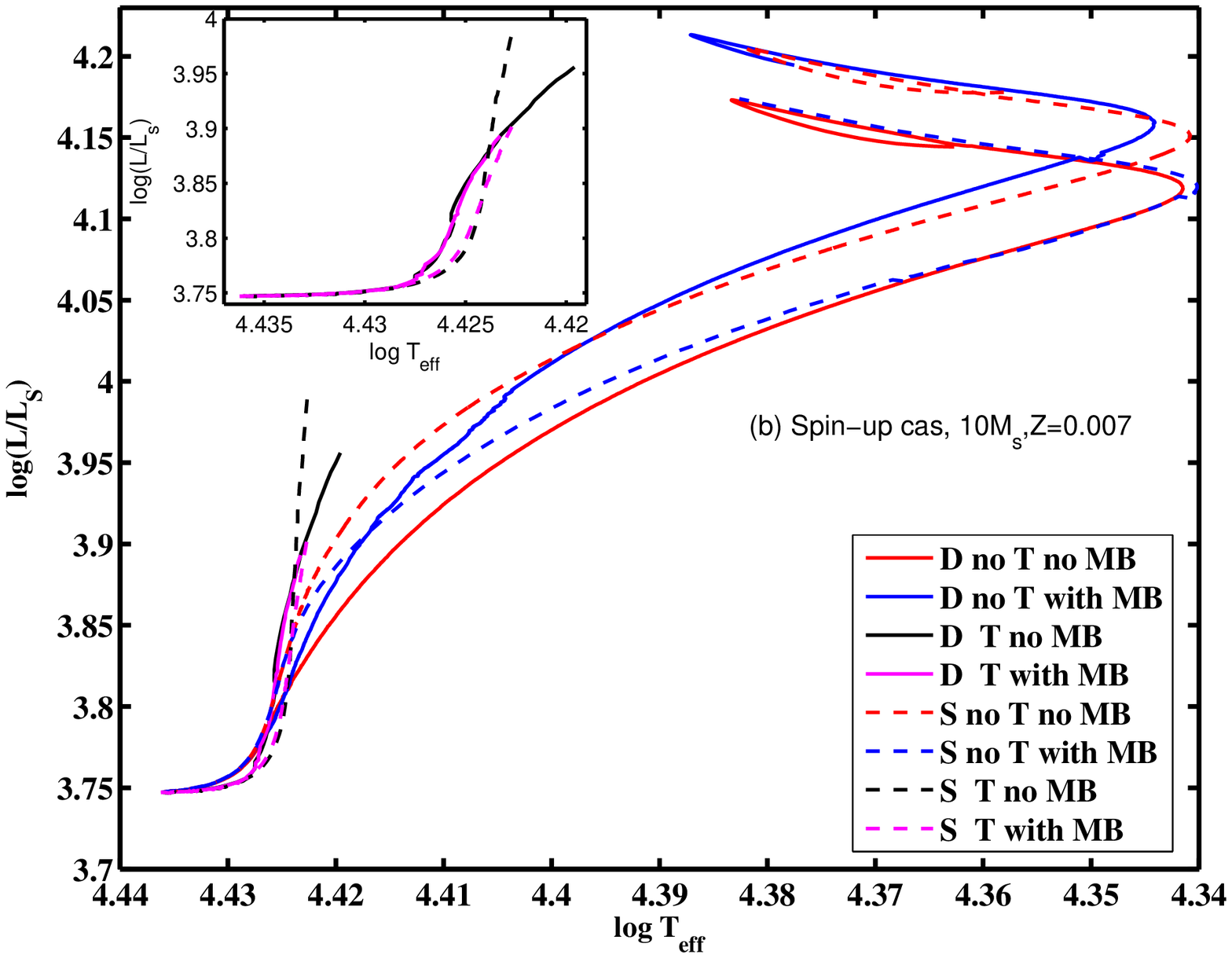}
   \caption{ Evolution in the theoretical HRD for the same models as shown in Fig.~\ref{v12}.
 {\it Left panel:} The star begins its evolution with a rotation on the ZAMS equal to 310 km s$^{-1}$. In the close binary considered, the star is being spun down by tidal forces.
   {\it Right panel:} The star begins its evolution with a rotation on the ZAMS equal to 60 km s$^{-1}$. In the close binary model considered, the star is spun up by tidal forces. The inset is a zoom of the beginning of the evolutionary tracks for the binary models.
   }
     \label{hr12}
  \end{figure*}

   \begin{figure*}
  \centering
  \includegraphics[width=8.9cm]{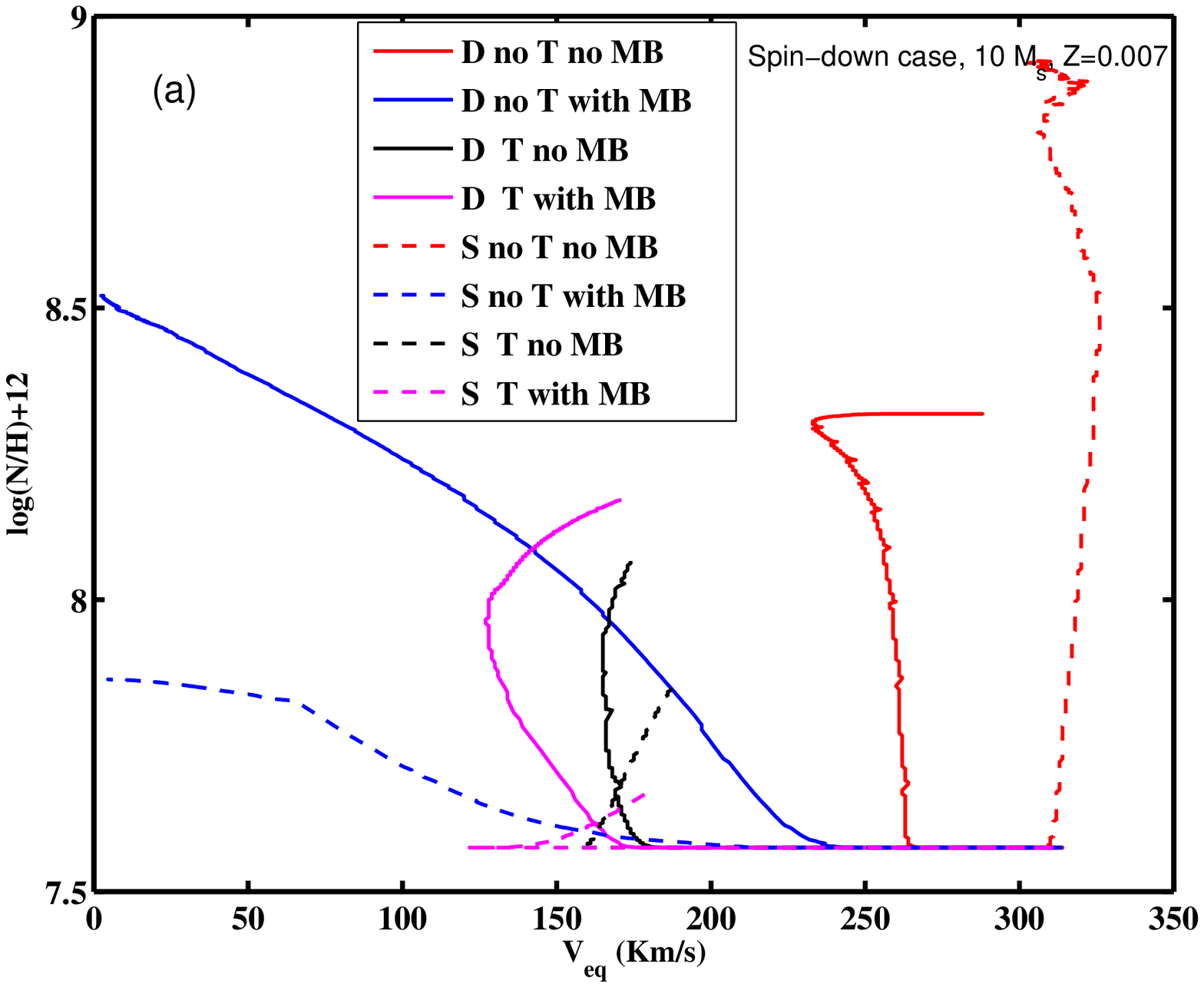}\hfill\hfill \includegraphics[width=8.9cm]{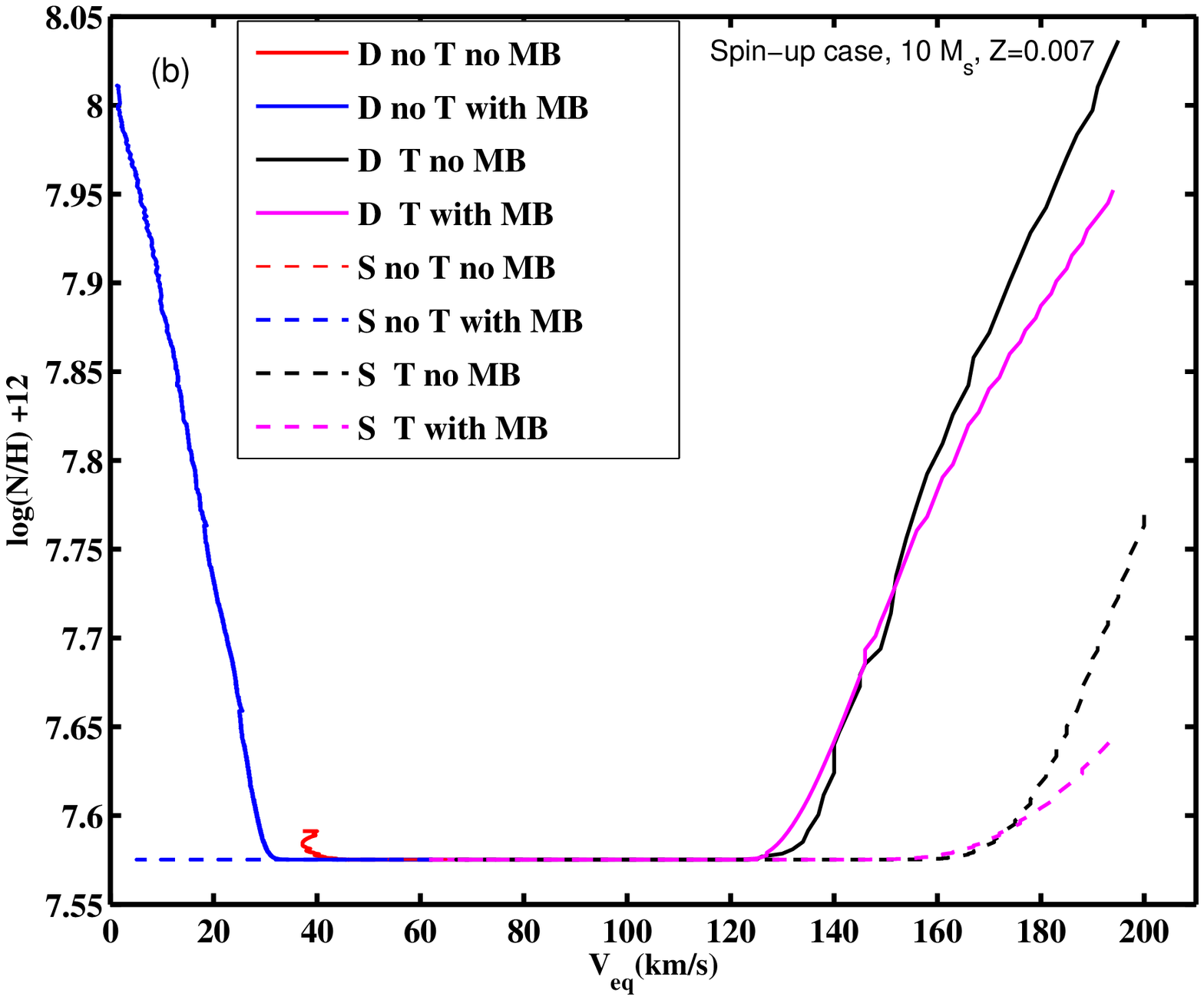}
   \caption{Evolution of the nitrogen abundance at the surface as a function of the surface velocity for the same models as those shown in Fig.~\ref{v12}. The abundances are expressed as ratios between the number of $^{14}$N atoms to the number of H atoms, normalized to a number of 10$^{12}$ H atoms.
{\it Left panel:} The star begins its evolution with a rotation on the ZAMS equal to 310 km s$^{-1}$. In the close binary considered, the star is spun down by tidal forces.
   {\it Right panel:} The star begins its evolution with a rotation on the ZAMS equal to 60 km s$^{-1}$. In the close binary considered, the star is spun up by tidal forces.
   }
     \label{n12}
  \end{figure*}

  \begin{figure*}
  \centering
  \includegraphics[width=8.9cm]{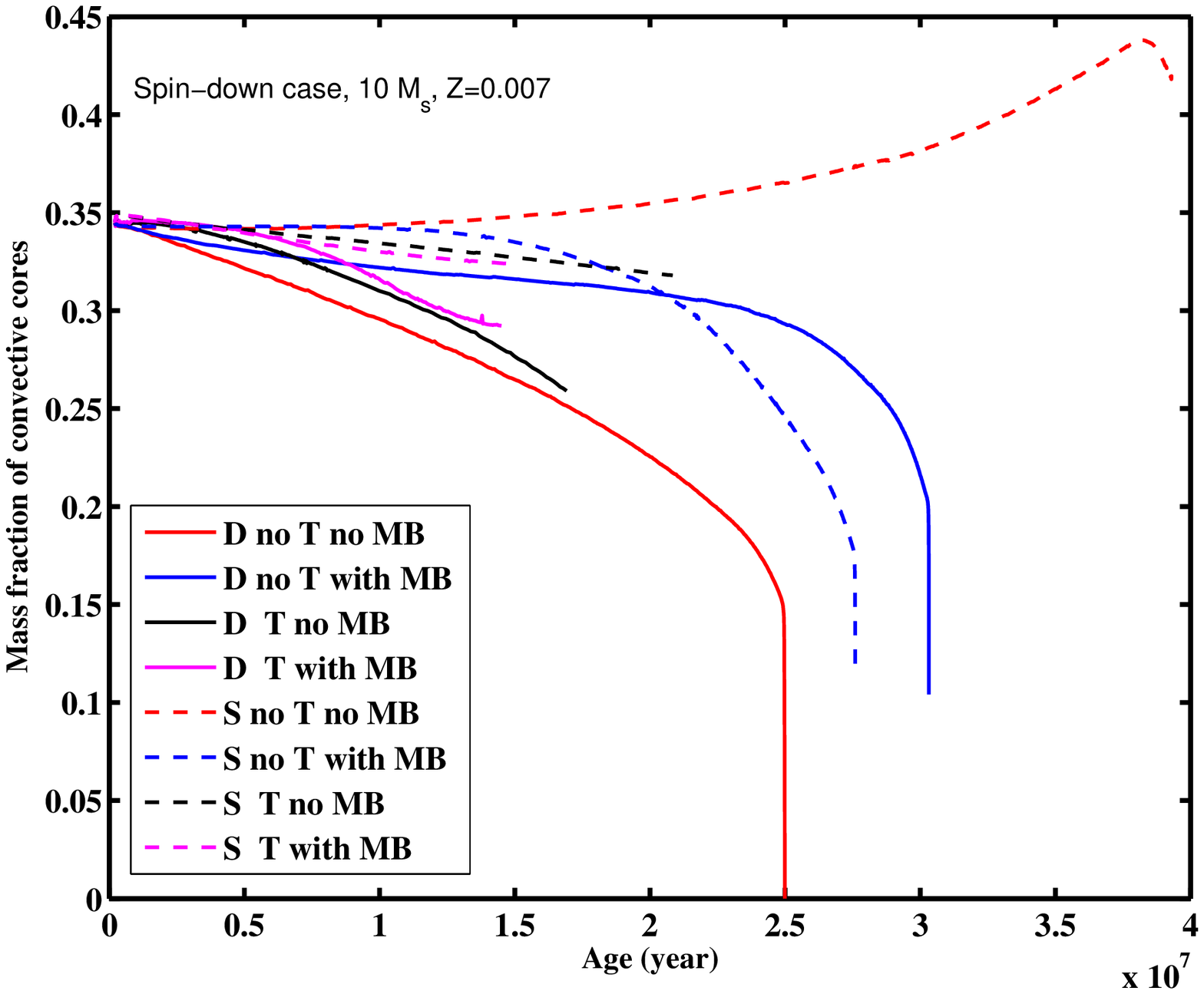} \includegraphics[width=8.9cm]{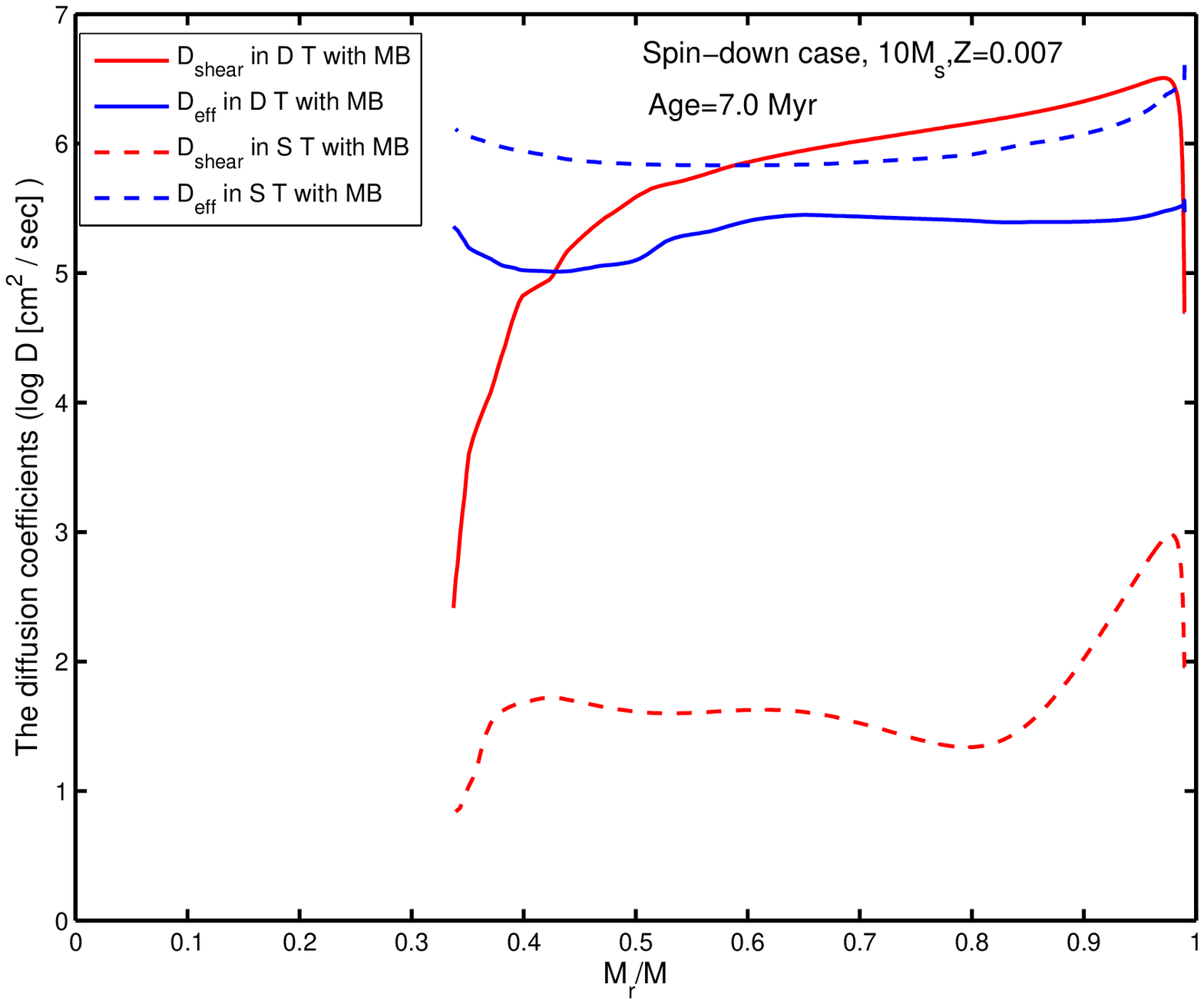}
   \caption{{\it Left:} Evolution of the mass fraction of the convective core as a function of time for the various models shown in the left panel of Fig.~\ref{n12}. {\it Right:} Variation as a function of the lagrangian mass (normalized to the total mass) of the shear ($D_{\rm shear}$)  and effective diffusion coefficients ($D_{\rm eff}$) in 10 M$_\odot$ models at an age of 7 Myr.}
     \label{tmc}
  \end{figure*}


  Figure~\ref{hr12} shows the evolutionary tracks for the same models as those shown in Fig.~\ref{v12}. The fast rotating models with no magnetic braking and no tidal forces (red curves) present quite different behaviors depending on
  the internal rotation profile. The solid-body rotating model follows a homogeneous evolution, while the differentially rotating model follows a more or less normal evolution (to the red part of the HR diagram). The initially slower-rotating models (see the red curves on the right panel) do not present any strong differences between differential and solid-body rotation. This is because
in both cases (solid-body and differentially rotating), the initial velocity considered is below the threshold for obtaining a homogeneous evolution. We just see that the solid-body rotating model is slightly more luminous because it undergoes a slightly more efficient mixing.

The non-interacting stars with magnetic braking (see the blue curves) present an interesting behavior: in cases of both high and moderate initial rotation, the models reaching the largest luminosities, that is, with
the strongest signs of mixing, are those with an internal differential rotation and no longer those with a solid body rotation, in contrast to models with no magnetic braking. This simply illustrates a result already discussed by \citet{Meynet2011},
showing that in a model with internal differential rotation, any braking at the surface reinforces the differential rotation and thus the shear mixing.
In a solid-body rotating star, the braking decreases $\Omega$ and thus the velocity of the
meridional currents, which are the main drivers for the mixing of chemical elements. Thus the solid-body rotating tracks for single stars with magnetic braking are less luminous and less strongly mixed.

The black  and magenta curves show the results when the star belongs to a close binary system.  We see that, depending on the internal angular momentum transport, the tracks are different. They are shifted to higher effective temperatures and luminosities when solid body rotation is accounted for. Thus, in close binaries the situation is qualitatively the same as for single stars.  This can be understood by the fact that tides cause these stars to evolve as if they have no braking. Actually, they are even slowly accelerated. This slow acceleration is not able to produce strong gradients in the outer layers and thus no additional mixing appears in the differentially rotating models.
There are few differences between the models with wind magnetic braking and those without. As already mentioned above, in close binaries tides dominate (at least for the conditions explored here).

This discussion indicates that the study of the evolutionary tracks in close binaries might be an interesting way to constrain the transport mechanisms induced by rotation. At a given age before mass transfer occurs, solid-body rotating models predict a larger luminosity-mass relation.
Let us note that considering a binary system allows us to somewhat constrain the age of the system in the sense that
a single isochrone should pass through the observed HR diagram positions of the two components of the binary system. Of course, the two components should have different initial masses.
In case eclipsing binaries are observed, then the study of the orbit yields the individual masses of the components
and thus tracks of the appropriate masses should pass through the two observed positions of the stars in the HR diagram at the same age. This makes the system tightly constrained. Cases with and without wind magnetic braking are predicted to
give very similar mass-luminosity relations, a characteristic that might be interesting to check as well. Finally, the situations for spin-down and spin-up are similar as well.

 \subsection{The evolution of the surface composition}

 The evolution of the surface enrichment in nitrogen is plotted as a function of the surface velocity in Fig.~\ref{n12} for the different models shown in Fig.~\ref{v12}\footnote{To simplify our calculations, we assume here that for those models with a surface magnetic field of fossil origin (i.e., not linked to a dynamo operating at the surface), the stronger magnetic field that is expected in the interior does not suppress either the shear mixing, or the meridional currents. Recent results suggest that this assumption may not be altogether accurate (e.g., Briquet et al. 2012).
}.


Let us first discuss the spin-down cases (left panel).
The single star models with fast rotation and no magnetic braking (red curves) evolve almost vertically in the nitrogen relative abundance versus surface rotation plot. The differentially rotating model shows the rapid initial decrease in velocity already discussed above. It also reaches a smaller nitrogen enrichment at the end of the MS phase.

The corresponding single star models with magnetic braking (blue curves) evolve toward slower rotation as expected. We see that before the end of the MS phase, the differentially rotating model reaches higher nitrogen surface enrichment than the corresponding model with no magnetic braking. As already discussed above, this results from the strong shear produced in the outer layers that triggers a stronger mixing compared to the non-magnetic model.
Single stars with wind magnetic braking and internal differential rotation might explain MS stars that are slow rotators and are highly nitrogen enriched, for example, the group 2 stars of \citet{Hunter2008}.
In the case of solid body rotation, we have
the reverse situation. The model with magnetic braking is much less enriched, because the braking process decreases $\Omega$, and decreasing $\Omega$ reduces the meridional currents that, in these models, are the main drivers of the
chemical mixing.

The close binary models evolve almost vertically, similar to the non-magnetic single star models. The differentially rotating models reach surface enrichments that are higher, at a given age, than the solid body-rotating cases.
This comes from the fact that torques applied at the surface of a star trigger stronger shear when the star rotates differentially. However,
in the previous section we saw that the solid-body rotating tracks had higher luminosities than the differentially rotating models, indicating a stronger degree of internal mixing.
How can we explain this apparent contradiction?
This apparent contradiction comes from the fact that evolution in the HRD depends on the distribution of the elements inside the entire star, and not only at the surface. The solid-body rotating models in binaries (black and magenta dashed curves in Fig.~\ref{tmc}) have significantly larger convective cores than in the analogous differentially rotating models (see the black and magenta continuous curves). Thus, globally, solid-body rotating stars are closer to a homogeneous structure than those that are differentially rotating.

Why are the convective cores larger and the surface nitrogen enrichments weaker in solid body rotating models compared to differentially rotating ones? 
To understand this, let us have a look at the right panel of Fig.~7. This plot shows the variation inside our 10 M$_\odot$ models (S and D models) of two diffusion coefficients, $D_{\rm shear}$ and  $D_{\rm eff}$ at an age of 7 Myr. The coefficient $D_{\rm shear}$ describes the diffusion due to the shear turbulence. The coefficient $D_{\rm eff}$ describes the mixing process accounting for the effects of a strong horizontal turbulence (responsible for the shellular rotation) and of the meridional currents \citep{Chaboyer1992}.

We see that in the solid body rotating model, the mixing of the elements is entirely due to the action of $D_{\rm eff}$. In the differentially rotating model, $D_{\rm eff}$ dominates
the mixing immediately above the convective core (between the mass coordinate 0.33-0.43) where the gradients of chemical composition reduce the shear turbulence, while $D_{\rm shear}$
dominates the transport in the outer parts of the radiative envelope. The fact that $D_{\rm eff}$ in the solid body rotating model is larger than in the differentially rotating one explains
why the convective core is larger in that model. The convective core in solid body rotating models is more efficiently replenished by fresh fuel present in the radiative envelope. On the other hand, the fact that
$D_{\rm shear}$ in the differentially rotating model is larger than $D_{\rm eff}$ in the solid body rotating case throughout the external radiative envelope explains why the surface enrichments
are larger in the differentially rotating model.

When lower initial rotation velocities are considered, we obtain the results shown in the right panel of Fig.~\ref{n12}.
All the single star models evolve to the left and all the close binary models evolve to the right.  The single star models with no magnetic braking show almost
no enrichment. Only the single star model with differential rotation and including wind magnetic braking shows a significant surface enrichment (by a factor of 2.5), although
the model began with a modest initial rotation, here around 65 km s$^{-1}$.
The corresponding solid-body rotating model shows no enrichment at all.

Differentially rotating close binary models present similar levels of enrichment to the single differentially rotating model with magnetic braking.
This is related to the fact that
a similar torque (although of different sign and of different origin) results from the wind magnetic braking and the tidal torque is produced during the slow equilibrium phase of the evolution of the surface velocity.
Similar torques produce similar gradients of $\Omega$ and thus trigger similar shear mixing.

The binary solid-body rotating analogs shows less enrichment than the differentially rotating models, although the enrichment is significantly larger than expected from single star models. The surface enrichment
in those models is mainly due to meridional currents.


From the evolution of the surface compositions, we conclude the same as for the HR diagram tracks.
The evolution of the nitrogen surface abundances of single stars is sensitive to  both the internal angular momentum transport and to the wind magnetic braking.
In close binaries, the nitrogen surface abundances are much more sensitive to the internal angular momentum transport than to the wind magnetic braking.

The variety of outputs obtained makes the interpretation of the positions of the stars in the nitrogen surface-enrichments versus surface-velocity plane rather complicated.
The above discussion shows that, for a given initial mass at a given initial metallicity, the position depends at least on the initial rotation, the age, the strength of the wind magnetic braking, the strength
of the tidal torques and the efficiency of the internal angular momentum transport.  The present numerical experiments show that few parts of that plane cannot be reached by considering some
special initial conditions. Population synthesis models accounting in a proper way for the distribution of the initial conditions for all the above quantities are needed to accurately understand whether
theory can provide a good fit of reality, at least in a statistical way.

\section{Conclusions and future perspectives}

We have studied the interactions between tides and wind magnetic braking in both differentially rotating and solid-body rotating
stellar models. We show that tides govern the initial phase of the evolution in these systems (at least for reasonable
magnetic fields and sufficiently short initial orbital periods). Then a near-equilibrium stage may set in, during which acceleration of the rotation by tides and deceleration by magnetic braking more or less compensate. This quasi-equilibrium may be reached after a short time and in that case it is maintained at least until
either the end of the MS phase or until a mass-transfer event occurs. Since during that phase angular momentum is continuously tapped from the orbit, the separation decreases.
As a consequence, this increases the accelerating force and thus increases the rotation rate of the star. Close binary evolution
may thus produce fast rotating, strongly magnetic main-sequence stars with ages that are well above the timescale for the wind magnetic braking.


The evolutionary tracks in the HR diagram of the close binaries do not depend significantly
on whether the star undergoes wind magnetic braking or not, but they are sensitive to the efficiency of the internal angular momentum transport.
We find that solid-body rotating models show higher luminosity-to-mass ratios than differentially rotating models.

Interestingly, we find that solid-body rotating models in close binaries are more globally mixed than analogous differentially rotating ones, but the surface enrichment in nitrogen
is larger in differentially rotating models. This difference between the degree of global and surface mixing emerges from
the different physics involved in the mixing of the chemical elements in differentially rotating and solid-body rotating models. In differentially rotating models, any change of $\Omega$ in the outer
layers reinforces the shear and thus the mixing in that region. For a similar velocity, solid-body rotation enhances the diffusion coefficient near the convective core, enlarging it.
This last effect makes the binary solid-body rotating models more luminous than the differentially rotating models. Of course, a larger domain of the parameter space should be explored.
This will be the subject of a forthcoming paper.

Observations of the mass-luminosity ratio, together with
nitrogen surface enrichment, of close binaries before the first mass transfer would provide interesting clues about
the physics of rotation. For those systems with a strong surface magnetic field, these observations would also
test whether or not a strong surface magnetic field can suppress the internal mixing of chemical species. As a consequence, confrontation of the predictions of these models using observations of real systems holds significant promise for advancing our understanding of stellar rotation, tidal interaction, and magnetic fields.

\begin{acknowledgements}
      This work was sponsored by the
Swiss National Science Foundation (project number 200020-172505), National Natural Science Foundation
of China (Grant No. 11463002), the Open Foundation of key Laboratory
for the Structure and evolution of Celestial Objects, Chinese
Academy of Science (Grant No. OP201405). GAW acknowledges Discovery Grant support from the Natural
Sciences and Engineering Research Council (NSERC) of Canada. T.F. acknowledges support from the
Ambizione Fellowship of the Swiss National Science Foundation (grant PZ00P2-148123).
\end{acknowledgements}


\bibliographystyle{aa}
\bibliography{song3}

\end{document}